\documentclass[aps,prb,footinbib,showpacs,showkeys,amssymb,amsmath,superscriptaddress,two column]{revtex4-1}
\pdfoutput=1
\usepackage{graphicx,graphics}

\usepackage[export]{adjustbox}


\usepackage{verbatim}
\usepackage{braket}
\usepackage{float}
\usepackage{breqn}
\usepackage{subfig}
\usepackage{color}
\usepackage{hyperref}
\usepackage{enumerate}

\makeatletter
\let\cat@comma@active\@empty
\makeatother

\begin{document}
	
	\title{Scaling of Quantum Geometry Near the Non-Hermitian Topological Phase Transitions}
	
	\author{Y R Kartik}
		\thanks{yrkartik@gmail.com}
	\affiliation{Institute of Atomic and Molecular Sciences, Academia Sinica, Taipei 10617, Taiwan}
 \author{ Jhih-Shih You}
		\thanks{jhihshihyou@ntnu.edu.tw}
	\affiliation{Department of Physics, National Taiwan Normal University, Taipei 11677, Taiwan}
  \author{H. H. Jen}
	\affiliation{Institute of Atomic and Molecular Sciences, Academia Sinica, Taipei 10617, Taiwan}
 \affiliation{Physics Division, National Center for Theoretical Sciences, Taipei 10617, Taiwan}
	 
	\date{\today} 
	
\begin{abstract}

The geometry of quantum states can be an indicator of criticality, yet it remains less explored under non-Hermitian topological conditions. In this work, we unveil diverse scalings of the quantum geometry over the ground state manifold close to different topological phase transitions in a non-Hermitian long-range extension of the Kitaev chain. The derivative of the geometric phase, as well as its scaling behavior,  shows that systems with different long-range couplings can belong to distinct universality classes. Near certain criticalities, we further find that the Wannier state correlation function associated with extended Berry connection of the ground state exhibits spatially anomalous behaviors. Finally, we analyze the scaling of the quantum geometric tensor near phase transitions across exceptional points, shedding light on the emergence of novel universality classes. 
\end{abstract}
	
\maketitle
\textit{Introduction-} 
Quantum geometry, which characterizes the geometrical structure of the eigenstate space, possesses a profound theoretical foundation~\cite{provost1980riemannian,torma2023essay,kolodrubetz2017geometry} and has been successfully measured across various experimental platforms~\cite{yu2020experimental, zheng2022measuring, yi2023extracting, cuerda2024observation,gianfrate2020measurement}. In particular, the connection or distance between adjacent quantum states in the parametric space could exhibit singular features close to quantum phase transitions~\cite{kolodrubetz2013classifying}. Therefore, quantum geometry plays a prominent role and aids in understanding critical behaviors, which is essential for determining physical properties, selecting appropriate working parameter spaces, and identifying quantum phase transitions~\cite{jiang2018topological, liao2021experimental, zhu2021band, zhang2019quantum, chen2024quantum}.

Non-Hermitian effects have brought significant advancements in quantum mechanics, particularly in areas such as topological phases of matter~\cite{ashida2020non}, unconventional criticality~\cite{arouca2020unconventional}, modified symmetry classifications~\cite{kawabata2019symmetry}, bulk-boundary correspondence~\cite{wu2022non,bao2021exploration}, scale-free localization~\cite{wang2022non,wang2023scaling}, the skin effect~\cite{xu2021coexistence,wang2022non}, higher-order exceptional points~\cite{mandal2021symmetry}, and geometry-dependent localization~\cite{zhou2023observation,zhang2022universal,wang2022non}. Furthermore, non-Hermitian effects in extended-range couplings have been investigated in relation to criticality~\cite{kumar2020multi,rahul2023unconventional}, non-Bloch band structures~\cite{bao2021exploration}, finite-temperature dynamics~\cite{kartik2023mixed}, and the emergence of exceptional points~\cite{mandal2021symmetry}. Recently, there have also been efforts to define quantum geometry in non-Hermitian systems~\cite{jiang2018topological, liao2021experimental, zhu2021band, zhang2019quantum, chen2024quantum}. However, little is known about the relation between the non-Hermitian criticality and the quantum geometry.

In this study, we investigate the scaling behaviors of the ground-state quantum geometry near topological phase transitions in one-dimensional non-Hermitian systems, focusing on a case study of the long-range Kitaev chain~\cite{kitaev2001unpaired,vodola2014kitaev}.  
The long-range Kitaev chain, which could be realized under various experimental platforms~\cite{li2013lattice,pientka2013topological,pientka2014unconventional,pientka2015topological,olekhno2022experimental,tao2020simulation,slim2024optomechanical}, exhibits unique phenomena~\cite{vodola2015long,viyuela2016topological,vodola2014kitaev, lepori2017long,cats2018staircase, kartik2021topological}.
The non-Hermiticity can be introduced into the model via imbalanced pair creation and annihilation. This can be implemented in open quantum systems where both the subsystem and the environment reside within a superconducting phase, characterized by weak interactions that facilitate particle-pair tunneling~\cite{li2018topological}.  
Notably, we reveal the emergence of novel universality classes through the derivative of the geometric phase and its scaling~\cite{cheng2017scaling,zhu2006scaling,nie2017scaling}, the Wannier state correlation behavior~\cite{chen2017correlation,chen2019topological}, and the quantum geometric tensor~\cite{jiang2018topological, liao2021experimental,zhu2021band, zhang2019quantum,chen2024quantum} over the ground state manifold. Finally, the adaptive nature of fidelity susceptibility enables us to understand critical behaviors near transitions happening through exceptional points.

\begin{figure}[H]
    \centering
        \includegraphics[width=\linewidth,height=10cm]{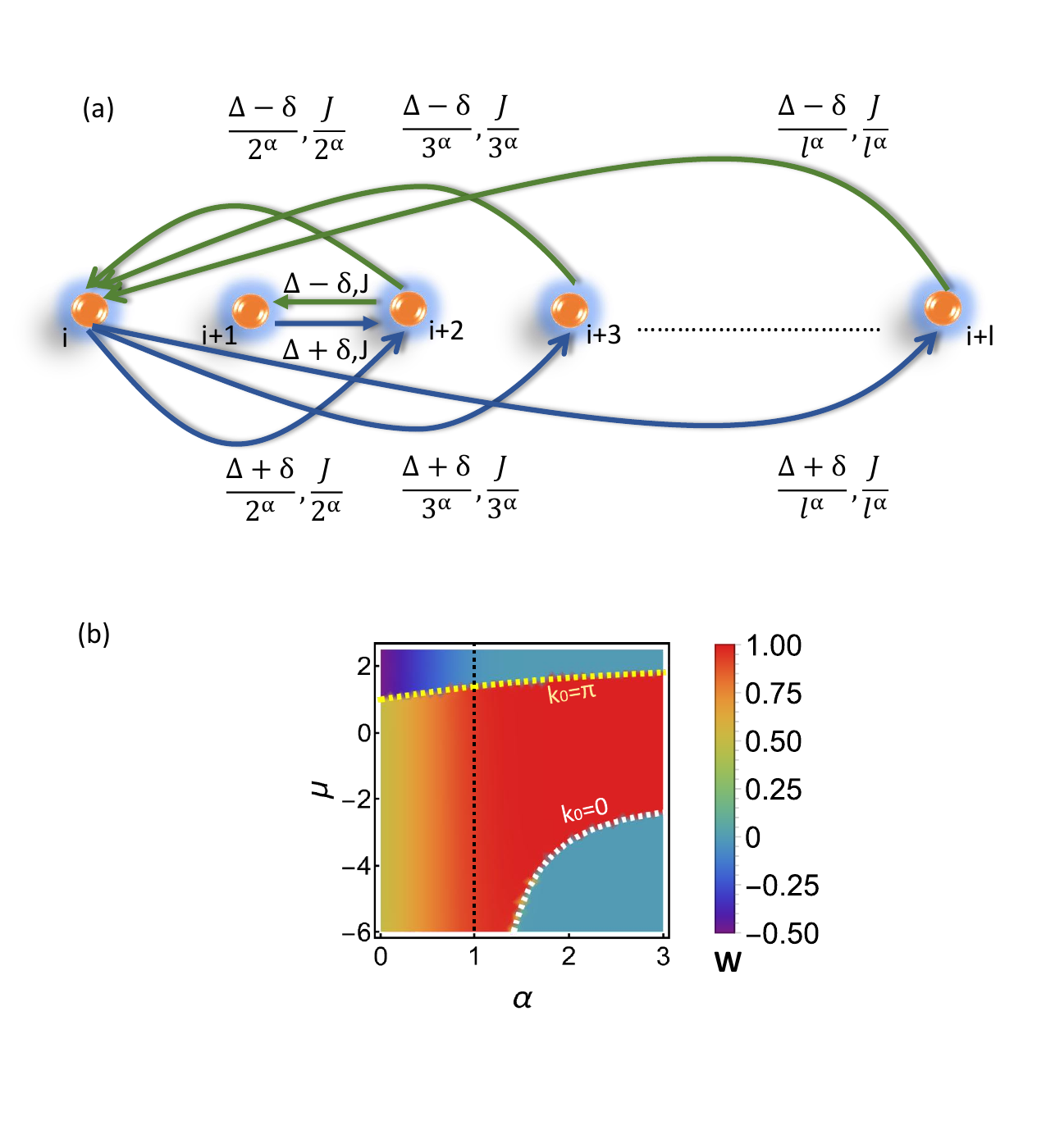}
    \caption{(Color online) (a) Schematic representation of the long-range non-Hermitian Kitaev chain as expressed in Eq.~\ref{e1}. (b) Phase diagram of the long-range non-Hermitian Kitaev chain in the \textit{PT}-unbroken region calculated through the extended Zak phase (Eq.~\ref{wn}) with \(\Delta=J=1\) and \(\delta=0.1\). Yellow and white lines represent analytical solutions \(\mu=-2J \text{Li}_{\alpha}(\mp 1)\) corresponding to \(k_0=\pi\) and \(k_0=0\) criticalities, respectively. The black line at \(\alpha=1\) represents the transition from the fractional to the topological phase occurring without gap closing.}
    \label{fig2}
\end{figure}

\textit{Model-}\label{Ham}
We consider a non-Hermitian Kitaev chain \cite{kitaev2001unpaired} with both imbalanced pairing \cite{li2018topological} and long-range couplings \cite{vodola2014kitaev,kartik2021topological}, as

\begin{eqnarray}
    H &=& \sum_{j=1}^{L} \mu c_j^{\dagger} c_j + \sum_{j=1}^{L-l} \sum_{l=1}^{r}  \frac{J}{l^{\alpha}} (c_j^{\dagger} c_{j+l} + c_{j+l}^{\dagger} c_j)\nonumber\\
    &+& \sum_{j=1}^{L-l} \sum_{l=1}^{r} \left(\frac{\Delta+\delta}{l^{\alpha}} c_j^{\dagger} c_{j+l}^{\dagger} + \frac{\Delta-\delta}{l^{\alpha}} c_{j+l} c_j \right),\label{e1}
\end{eqnarray}

\noindent where \(c_j (c_j^{\dagger})\) denotes the annihilation (creation) operator, and \(r\) represents the range of neighboring couplings, which extends to infinity (Fig.~\ref{fig2} a). The parameters \(J\), \(\Delta\), \(\delta\), and \(\mu\) are real quantities corresponding to hopping, pairing, non-Hermitian imbalance, and chemical potential, respectively. Long-range effects are implemented via a power-law decay of both hopping and pairing amplitudes with lattice distance, characterized by decay exponent \(\alpha\). In the limit \(\delta = 0\) and \(\alpha \rightarrow \infty\), this model reduces to the Hermitian short-range case. The non-Hermitian model, which respects time-reversal symmetry, particle-hole symmetry, and chiral symmetry but lacks sub-lattice symmetry, falls into the BDI class of the real Altland-Zirnbauer~(AZ) symmetry classification~\cite{li2018topological,kawabata2019symmetry,supp}. It is also evident that the model obeys an additional symmetry called the parity-time~(\textit{PT}) symmetry. 

The long-range Hamiltonian can also be expressed in the Bogoliubov–de Gennes~(BdG) form as~(Supplementary Material~\cite{supp})
\begin{equation}
H_{\text{BdG}} = 
\begin{pmatrix}
-\mu - 2J f_{\alpha}(k) & -2i (\Delta+\delta) g_{\alpha}(k) \\
2i (\Delta-\delta) g_{\alpha}(k) & \mu + 2J f_{\alpha}(k)
\end{pmatrix},
\end{equation}
where
\begin{eqnarray}
f_{\alpha}(k) &=& \sum_{l=1}^{L-r} \frac{\cos(kl)}{l^{\alpha}} = \frac{\text{Li}_{\alpha}(e^{ik}) + \text{Li}_{\alpha}(e^{-ik})}{2},\nonumber\\
g_{\alpha}(k) &=& \sum_{l=1}^{L-r} \frac{\sin(kl)}{l^{\alpha}} = \frac{\text{Li}_{\alpha}(e^{ik}) - \text{Li}_{\alpha}(e^{-ik})}{2i},
\end{eqnarray}
and \(\text{Li}_{\alpha}\) is the polylogarithmic function~\cite{vodola2014kitaev}. 
Diagonalizing $H_{\text{BdG}}$ gives the quasi-energy spectrum,
\begin{eqnarray}
E_k &=& \sqrt{(-\mu - 2J f_{\alpha}(k))^2 + 4(\Delta^2 - \delta^2) g^2_{\alpha}(k)}.\label{ed}
\end{eqnarray}
Therefore, we have $H = \sum_k E_k \bar{A}_k A_k$ with the ground state $|g\rangle\rangle$ or $|\bar{g}\rangle$ defined by $A_k |g\rangle\rangle =0$ or $\bar{A}^{\dagger}_k |\bar{g}\rangle =0,$ where $A_k$ and $\bar{A}_k$ are complex Bogoliubov
modes fulfilling the anti-commutation relations $\{ A_k, \bar{A}_{k'} \} = \delta_{k,k'}$
and $ \{ A_k, A_{k'} \} = \{ \bar{A}_k, \bar{A}_{k'} \} = 0$~\cite{supp}. The Hamiltonian attains pseudo-Hermiticity, but the imaginary spectrum can appear for the condition $(-\mu - 2J f_{\alpha}(k))^2 + 4(\Delta^2 - \delta^2) g^2_{\alpha}(k)<0$, which corresponds to the spontaneously \textit{PT}-broken region. 
In the symmetry-unbroken region, $H_{\text{BdG}}$ can be mapped to its Hermitian counterpart through a surrogate Hamiltonian as~\cite{li2018topological}
\begin{equation}\label{surr}
H_{\text{s}}(k) = \begin{pmatrix}
-\mu - 2J f_{\alpha}(k) & i \sqrt{\Delta^2-\delta^2} g_{\alpha}(k) \\
-i \sqrt{\Delta^2-\delta^2} g_{\alpha}(k) & \mu + 2J f_{\alpha}(k)
\end{pmatrix},
\end{equation}
whose quasi-energy spectrum is fully real.

The topological property of the quasi-energy spectrum can be characterized by the extended Zak phase, defined as~\cite{supp}
\begin{eqnarray}\label{wn}
W &=&\frac{1}{2\pi} \int_{-\pi}^{\pi} \langle\psi  | \partial_k | \psi\rangle \rangle  dk, 
\end{eqnarray}
\noindent where \( | \psi \rangle \) and \( | \psi \rangle\rangle \) are biorthonormal eigenstates of $H^{\dagger}_{\text{BdG}}$ and $H_{\text{BdG}},$ respectively. 
This expression generally holds in all parameter spaces except at degeneracy points.

With the extended Zak phase, we can first establish the topological phase diagram as a function of decay exponent $\alpha$ and chemical potential $\mu$ in the \textit{PT} symmetry-unbroken region as shown in Fig.~\ref{fig2}~(b). At $\alpha = 1$ the fractional and the integer Zak phases are continuously connected without gap closing in the spectrum. At \(\mu_c=-2J \text{Li}_{\alpha}(+ 1)\) and \(-2J \text{Li}_{\alpha}(- 1)\), we find that the degenerate ground states with gapless spectra \(E_k=0\) occur at \(k_0=0\) and \(\pi\), respectively. Thus, each $\mu_c$ and corresponding $k_0$ serves as a criticality separating two Zak phases with an integer difference.
Under criticalities we start by investigating the scaling of the gapless spectrum, expressed as \(E_k\propto k^z\), with \(z\) as the dynamical critical exponent. We observe that, for a given \(\Delta\), the \(k_0=\pi\) criticality is sensitive to the non-Hermitian parameter \(\delta\) and remains unaltered with the variation of \(\alpha\), where the spectrum is linear~(\(z=1\)) for \(\Delta>\delta\), quadratic~(\(z=2\)) for \(\Delta=\delta\), and root-square dispersive~(\(z=1/2\)) for \(\Delta<\delta\), respectively. On the other hand, the \(k_0=0\) criticality is sensitive to the long-range parameter~\(\alpha\) and remains unchanged for a given \(\Delta\) and \(\delta\), where the spectrum is linear~(\(z=1\)) for \(\alpha>2\), fractional~(\(z=1/4\)) for \(1<\alpha<2\), and gapped for \(\alpha<1\), respectively (Supplementary Material~\cite{supp}).

It was revealed that degeneracy can also be captured by the local geometry of the ground state~\cite{zhu2006scaling}. Next we will analyze the behavior of the geometric phase and corresponding critical exponents around the criticalities.

\begin{figure}[H]
\centering
\includegraphics[width=\columnwidth,height=6cm]{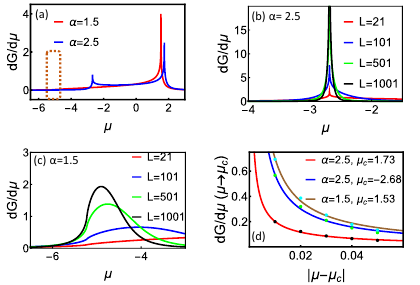}
\caption{(Color online) (a) First-order derivative of the geometric phase.
For \(\alpha > 2\), around criticalities corresponding to both \(k_0 = 0\) and \(k = \pi\), $\frac{dG}{d\mu}$ exhibits non-analytic peaks. For \(1 < \alpha < 2\), while a non-analytic peak still appears near the  \(k_0 = \pi\) criticality,  $\frac{dG}{d\mu}$ fails to display such a non-analytic behavior near the  \(k_0 = 0\) criticality, as shown in the brown box. (b) Near the $k_0=0$ criticality in the range $\alpha > 2$, non-analytic peaks increase with system size. (c) Near the $k_0=0$ criticality in the range $1 < \alpha < 2$, $\frac{dG}{d\mu}$ does not exhibit non-analyticity. (d) The non-analytic peaks in (a) are associated with critical exponent \(\nu \approx 1\), with parameters \(J = \Delta = 1\), \(\delta = 0.2\).}
\label{gps}
\end{figure}

\textit{Geometric Phase Scaling-} \label{GPS}
  The geometric phase of the ground-state wavefunction $ |g\rangle\rangle$ is~\cite{cheng2017scaling,zhu2006scaling,nie2017scaling}
\begin{equation}
G = - \frac{\pi}{L}\sum_{k} (1 - \cos \theta_k),
\end{equation}
where 
\begin{equation}
  \cos \theta_k = \frac{-\mu - 2J f_{\alpha}(k)}{\sqrt{(-\mu - 2J f_{\alpha}(k))^2 + (2\sqrt{\Delta^2 - \delta^2} g_{\alpha}(k))^2}}.  \nonumber
\end{equation}
In the thermodynamic limit, as \(L \rightarrow \infty\), the summation \(\frac{1}{L}\sum_{k}\) can be approximated by the integral \(\frac{1}{2\pi}\int_{-\pi}^{\pi} d\phi\) with \(\phi = \frac{2\pi k}{L}\).  
In general, the derivative of the geometric phase exhibits a divergence in the vicinity of the 
criticality~\cite{zhu2006scaling}. This should imply a topological phase transition in our non-Hermitian system. Therefore, we plot the derivative of the geometric phase against the chemical potential for different values of $\alpha$ in Fig.~\ref{gps}(a). First we note that for \(\alpha > 2\), around criticalities $ \mu_c=-2J\text{Li}_{\alpha}(1)$ and $-2J\text{Li}_{\alpha}(-1),$ which correspond to $k_0 = 0$ and to \(k_0 = \pi\) respectively, $\frac{dG}{d\mu}$ indeed exhibits non-analytic peaks; however for \(1 < \alpha < 2\), only around the \(k = \pi\) criticality $\frac{dG}{d\mu}$ exhibits a non-analytic peak.

The derivatives of geometric phase are also expected to exhibit small peaks for smaller system sizes, gradually shifting toward criticality and becoming non-analytic as the system size increases, following a scaling relation $\frac{dG}{d\mu}\big|_{\mu_m} \propto \ln(N)$, where $\mu_m$ is the position of peaks and referred to as pseudo-criticalities~\cite{zhu2006scaling}. At \(\mu_m\) corresponding to \(k_0 = 0\), the $\frac{dG}{d\mu}$ shows increasing spikes with system size, signaling non-analytic behaviors for $\alpha > 2$ ~(Fig.~\ref{gps}b). For $1 < \alpha < 2$, however, with increasing system size it does not show a non-analytic behavior~(Fig.~\ref{gps}c). 

\indent The non-analytic peaks in the thermodynamic limit exhibit a scaling behavior
\begin{equation}
    \frac{dG}{d\mu} \propto \frac{1}{1 + \xi},
\end{equation}
where the characteristic length $\xi \propto |\mu - \mu_c|^{-\nu}$, with $\mu_c$ as the critical value and $\nu$ as the critical exponent.
Near $ \mu_c=-2J\text{Li}_{\alpha}(\pm1)$ for $\alpha > 2$ and near $ \mu_c=-2J\text{Li}_{\alpha}(-1)$ for $1 < \alpha < 2$ we numerically find the critical exponent $\nu \approx 1$, as shown in Fig.~\ref{gps}(d).
Consequently, our results related to the derivative of the geometric phase signify the emergence of a distinct universality class, where the criticality exists but the critical exponents cannot be defined for a certain range of parameters. 

To further understand the distinct universality classes, we will investigate spatial behaviors and scaling properties of the Wannier state correlation function associated with extended Berry connection.

\textit{Wannier State Correlation Function-}  
For a two-band model in 1D, the Wannier state correlation function acts as the real-space position correlation between filled-band Wannier states located at different positions~\cite{chen2017correlation,malard2020scaling,malard2020multicriticality,abdulla2020curvature,molignini2020generating,chen2018weakly,chen2016scaling,kumar2020multi,kumar2020quantum,kumar2023topological,kumar2023signatures}. Here we consider the Wannier-state representation of a lower eigenstate of $H_{\text{BdG}}$ positioned at $R$ as~\cite{chen2017correlation,kumar2020multi,kumar2020quantum},
\begin{equation}
    |R\rangle=\int e^{ik(\hat{r}-R)}|\psi\rangle dk,
\end{equation}
with $\hat{r}$ as the position operator. Therefore, the position correlation between Wannier states centered at $|0\rangle\rangle$ and $|R\rangle$ is
\begin{equation}
    \lambda_R=\langle R|\hat{r}|0\rangle\rangle=\frac{1}{2\pi}\int  e^{ikR} \langle \psi |i \partial_k | \psi\rangle \rangle dk,
\end{equation}
where  $\langle \psi |i \partial_k | \psi\rangle \rangle$ can be written as $\langle g |i \partial_k |g\rangle \rangle,$ which corresponds to the extended Berry connection of the ground state.
\begin{figure}[H]
	\centering
	\includegraphics[width=\columnwidth,height=3.4cm]{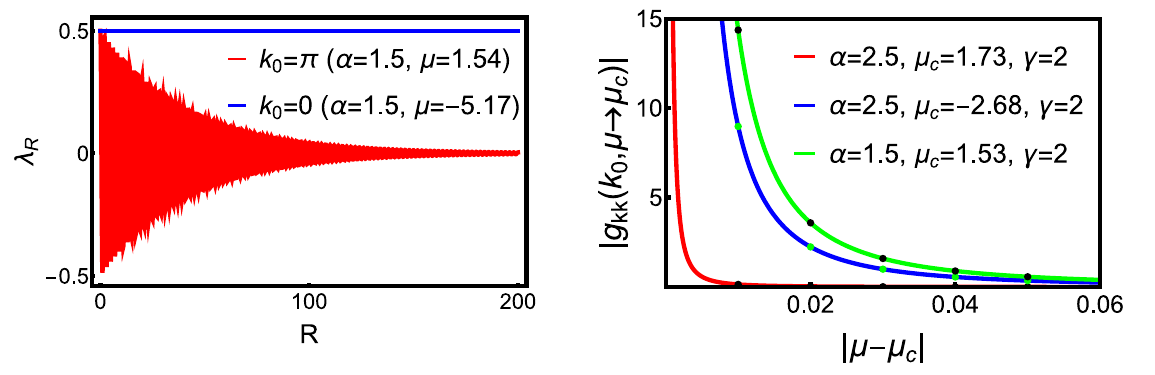}
	\caption{(Color online) (a) Spatial behavior of the Wannier correlation function for \(\alpha = 1.5\) and within the integer Zak phase. Here we choose $\mu=1.54$ which is near the $k_0 = \pi$~($\mu_c=1.53$) criticality and $\mu=5.17$ which is near $k_0 = 0$~($\mu_c=5.22$) criticality. (b) Scaling of fidelity susceptibility yielding a critical exponent, which is absent around $k_0=0$ for the range $1<\alpha<2$, and gives \(\gamma = 2\) around other criticalities. }
	\label{f3}
\end{figure}
Near $k_0$ criticality, the Wannier correlation function, which can be calculated from the Fourier transform of the extended Berry connection in the Ornstein-Zernike form, yields
\begin{eqnarray}
    \lambda_R &=& \oint \frac {dk}{2\pi} \langle \psi | i\partial_k | \psi\rangle \rangle e^{ikR}\nonumber\\
    &\approx&e^{ik_0R}\frac{\langle \psi | i\partial_k | \psi\rangle \rangle|_{k_0}}{2\xi}e^{\frac{-|R|}{\xi}},
\end{eqnarray}
where $\xi$ is the correlation length behaving as  $\xi\propto|\mu-\mu_c|^{-\nu}$.  
In our model for $1<\alpha<2$ and within the integer Zak phase,  $\lambda_R$ close to $k_0=\pi$ criticality exponentially decays, while close to $k_0=0$ criticality, it has a constant amplitude throughout the sites~(Fig.~\ref{f3}a).  The Wannier correlation function shows a spatially anomalous behavior close to criticalities, even though their topological characteristics remain the same.

\textit{Quantum Geometric Tensor-} \label{Scale}
The quantum geometric tensor measures the geometric distance between proximate quantum states, which is a powerful tool for analyzing quantum phase transitions~\cite{provost1980riemannian,kolodrubetz2013classifying,torma2023essay}. For non-Hermitian systems, the quantum geometric tensor, which is gauge-invariant, can be defined as~\cite{jiang2018topological, liao2021experimental, zhu2021band, zhang2019quantum, chen2024quantum}
\begin{eqnarray}
    \chi_{aa^{\prime}} = \frac{1}{2}\left[ \langle \partial_{a} \psi | \partial_{a^{\prime}} \psi\rangle \rangle - \langle \partial_{a} \psi | \psi\rangle \rangle \langle \psi | \partial_{a^{\prime}} \psi\rangle \rangle \right] 
   + \frac{1}{2}\left[a \Leftrightarrow a^{\prime}\right],\nonumber
\end{eqnarray}
where $a,a^{\prime}$ represent the parameters of the system and the biorthonormal state $|\psi\rangle$ and $|\psi\rangle\rangle$ in this study are considered to be the ground state wavefunction.
Here, the term $a \Leftrightarrow a^{\prime}$ ensures the symmetric nature of the quantum geometric tensor over parameter exchange. Its real and imaginary components correspond to fidelity susceptibility and Berry curvature, respectively:
\begin{eqnarray}
g_{aa^{\prime}} &=& \text{Re}[\chi_{aa^{\prime}}] = \frac{1}{2}(\chi_{aa^{\prime}} + \chi_{a^{\prime}a}), \nonumber \\
F_{aa^{\prime}} &=& \frac{1}{2}i(\chi_{aa^{\prime}} - \chi_{a^{\prime}a}) = \partial_{a} A_{a^{\prime}} - \partial_{a^{\prime}} A_{a},
\end{eqnarray}
\noindent with \( A_{a} \) corresponding to the extended Berry connection over the ground state manifold. 
\begin{figure}[H]
	\centering
	\includegraphics[width=\columnwidth,height=8cm]{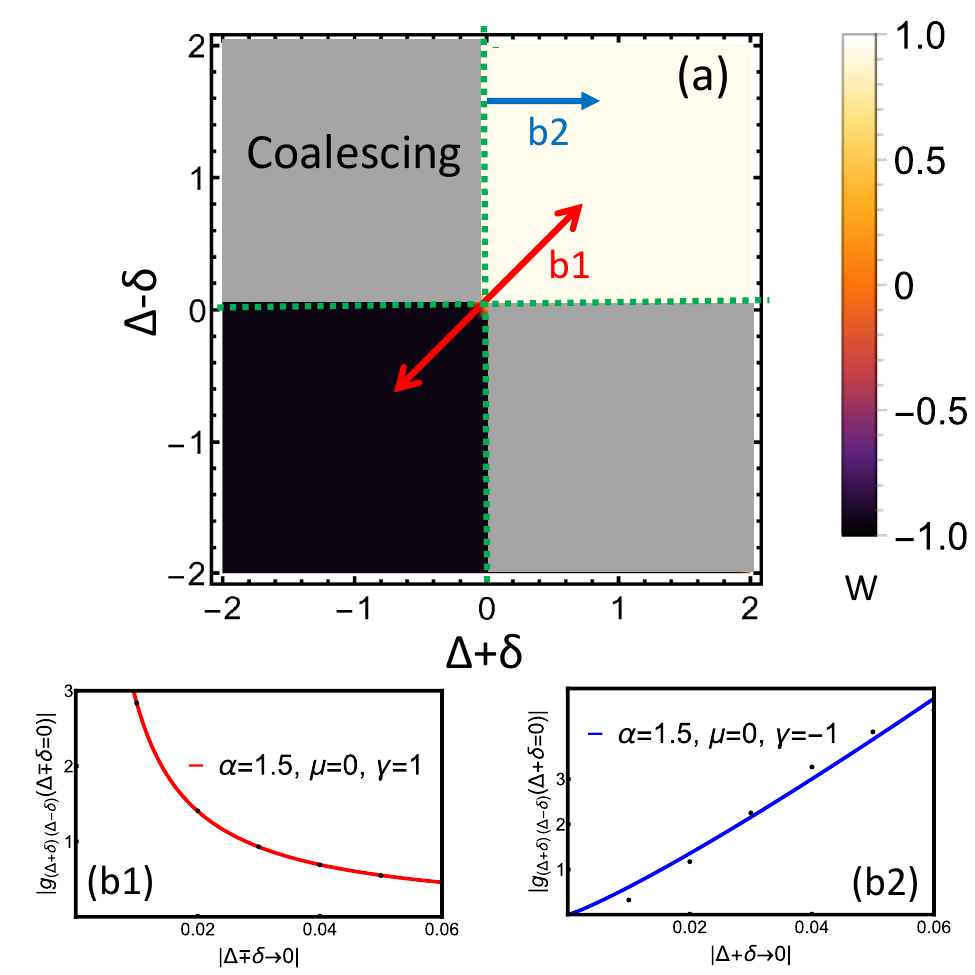}
	\caption{(Color online) (a) Phase diagram for imbalanced pairings with parameter $J=1,\alpha=1.5,\mu=0.$  Topological phases ($W=\pm1$) for $\Delta^2-\delta^2>0$ and coalescing phases for $\Delta^2-\delta^2<0$. (b1) Scaling of ground state fidelity susceptibility $g_{(\Delta+\delta)(\Delta-\delta)}$ with critical exponent \(\gamma = 1\) for the transition between $W=1$ and $W=-1$ around $\Delta=\delta=0$~(denoted by red arrow). (b2) Scaling of ground state fidelity susceptibility $g_{(\Delta+\delta)(\Delta-\delta)}$ with critical exponent \(\gamma \approx -1\) for the transition between $W=1$ and coalescing phase~(denoted by blue arrow). }
	\label{f32}
\end{figure} 
First we examine the fidelity susceptibility of the ground state with respect to quasi-momentum~$k$, i.e., $g_{kk},$ near $k_0$ criticality. We find that $g_{kk}$ scales as the squared extended Berry connection under biorthonormal bases~\cite{supp} and satisfies the scaling relation:
\begin{eqnarray}
g_{kk} \big|_{k=k_0} \propto |\mu - \mu_c|^{-\gamma},
\end{eqnarray}
where $\gamma$ is the critical exponent.
Fig.~\ref{f3}(b) shows that around $k_0=0,\pi$ for \(\alpha>2 \) and around $k_0=\pi$ for \(1<\alpha < 2\), the fidelity susceptibility becomes non-analytic, and upon scaling, it yields a critical exponent \(\gamma = 2\). For \(1<\alpha < 2\) around \(k_0 = 0\), however, the fidelity susceptibility does not follow a scaling behavior, resulting in an undefined critical exponent.


The adaptive nature of fidelity susceptibility allows itself to extend towards different parameter spaces, helping us understand different critical behaviors. Therefore, the scaling theory of the ground state fidelity susceptibility can not only be efficiently applied around the criticalities associated with \textit{PT}-symmetry unbroken regions, but also to phase transitions across exceptional points, which separate \textit{PT-}unbroken and \textit{PT-}broken regions. The exceptional points occur in the parameter space when the complex eigenvalues and corresponding eigenvectors of the $H_{\text{BdG}}$ become coalescing. As a result, such non-Hermitian degeneracies lead to the formation of a single self-orthogonal ground state.       

The long-range Kitaev model with imbalanced pairings can host integer Zak phases for $\Delta^2-\delta^2>0$ where the \textit{PT} symmetry of states is preserved, and coalescing phases for $\Delta^2-\delta^2<0$ with spontaneous \textit{PT} symmetry breaking, as shown in Fig.~\ref{f32}a. 
We observe that fidelity susceptibility gives a finite value for $\Delta^2-\delta^2>0$ and becomes divergent due to exceptional points in the coalescing region, exactly reproducing the phase diagram under the given parameter space. Fig.~\ref{f32}a shows the transition between $W=1$ and $W=-1$ at $\Delta=\delta=0$, where the $H_{\text{BdG}}$ reduces to a null matrix~\cite{li2018topological}. We find that the fidelity susceptibility of the ground state, $g_{(\Delta+\delta)(\Delta-\delta)},$ captures the scaling around $\Delta=\delta=0$ with a critical exponent $\gamma=1$~(Fig.~\ref{f32}b1).

The phase boundaries between topological ($W = \pm1$) and coalescing phase shown in Fig.~\ref{f32}a are exceptional points. Since the fidelity susceptibility remains divergent throughout the coalescing phases, determining the scaling behavior within the phases is not accessible. Nevertheless, for the transition between topological and coalescing phases, we can still perform a scaling analysis for the fidelity susceptibility near the boundary from the topological phase side. This yields a critical exponent $\gamma = -1$~(Fig.~\ref{f32}b2).\\

\indent \textit{Conclusion-} In this work, we investigate the scaling behaviors of the ground-state quantum geometry near various topological phase transitions in the presence of non-Hermiticity. For this, we consider a non-Hermitian Kitaev chain with long-range hopping labeled by decay exponent $\alpha$. In the \textit{PT} symmetry-unbroken region, we observe that, although the extended Zak phase is quantized for \(\alpha > 1\), the derivative of the geometric phase, as well as its scaling behavior, shows a distinct universality class for the range \(1<\alpha<2\) around the \(k_0 = 0\) criticality, where the corresponding Wannier state correlation function also exhibits a spatially non-decaying behavior. The ground-state fidelity susceptibility enables us to disclose the critical scaling around exceptional transition points which separate \textit{PT-}unbroken and \textit{PT-}broken regions. Our work paves the way towards future research on the quantum geometry and its scaling near non-Hermitian topological phase transitions in other exotic topological and spin systems.

\textit{Acknowledgments-} H. H. J. acknowledges support from the National Science and Technology Council (NSTC), Taiwan, under Grant No. NSTC-112-2119-M-001-007, No. NSTC-112-2119-M-001-079-MY3, and from Academia Sinica under Grant AS-CDA-113-M04. J.-S.Y. acknowledges support from the National Science and Technology Council (NSTC), Taiwan, under Grant No. NSTC 113-2112-M-003-015 -.We also express our gratitude for support from TG 1.2 and TG 3.2 of NCTS at NTU.

\newpage
\section*{Supplementary Material}
\section{Complex Bogoliubov Transformation}\label{appb}

After performing a Fourier transformation, Eq. 1 (in the main text) can be expressed in the momentum basis as:
\begin{equation}
    c_j = \frac{1}{\sqrt{N}} \sum_k e^{ikj} c_k,
\end{equation}
where \(k = \frac{2\pi m}{N}, \hspace{0.2cm} (m = 0, 1, 2, \ldots, N-1)\). Consequently, we obtain:
\begin{eqnarray}
    H &=& \sum_k (-2Jf_{\alpha} - \mu) c_k^{\dagger} c_k \\
    &+& \sum_k \left[ ig_{\alpha} \left( (\Delta - \delta)c_{-k}c_k + (\Delta + \delta)c_{-k}^{\dagger} c_k^{\dagger} \right) \right]. \nonumber
\end{eqnarray}

Due to the asymmetry in the forward-backward coupling of the pairing operator, a complex Bogoliubov transformation is employed to achieve the Bogoliubov-de Gennes (BdG) form. In this context, the BdG Hamiltonian represents a superposition of electron and hole states, manifesting fermionic behavior in a complex formulation:

\begin{eqnarray}
    A_k &=& u_k c_k - v_k c_{-k}^{\dagger}, \nonumber \\
    \bar{A}_k &=&\bar{u}_k c^{\dagger}_k - v_k c_{-k}.\label{qp}
\end{eqnarray}
with $v^2_k+u_k\bar{u}_k=1.$

These complex Bogoliubov modes fulfill the anti-commutation relations:
\begin{eqnarray}
    \{ A_k, \bar{A}_{k'} \} &=& \delta_{k,k'}, \nonumber \\
    \{ A_k, A_{k'} \} &=& \{ \bar{A}_k, \bar{A}_{k'} \} = 0,
\end{eqnarray}
which diagonalize the Hamiltonian as follows:
\begin{equation}
    H = \sum_k E_k \bar{A}_k A_k,
\end{equation}
with \(E_k=\sqrt{(-\mu - 2J f_{\alpha}(k))^2 + 4(\Delta^2 - \delta^2) g^2_{\alpha}(k)}\) as specified in the main text. The Hamiltonian is non-Hermitian, as \(\bar{A}_k \neq A_k^{\dagger}\). When the ground state can be defined by $A_k |g\rangle\rangle =0$ or $\bar{A}^{\dagger}_k |\bar{g}\rangle =0,$ the right and left eigenstates of \(H\) can be expressed as:
\begin{equation}
    \Pi_{\{ k \}} \bar{A}_k |g\rangle\rangle \quad \text{and} \quad \langle \bar{g}| \Pi_{\{ k \}} A_k,
\end{equation}
where \(|g\rangle\rangle\) and \(| \bar{g}\rangle\) denote the ground states of \(H\) and \(H^{\dagger}\), respectively. Here, \(\bar{A}_k\) and \(A_k\) take the following form:
\begin{eqnarray}
    A_k &=& i\sqrt{\frac{\Delta-\delta}{\Delta+\delta}} \xi_k c_k + \eta_k c_{-k}^{\dagger}, \nonumber \\
    \bar{A}_k &=& -i\sqrt{\frac{\Delta+\delta}{\Delta-\delta}} \xi_k c_k^{\dagger} + \eta_k c_{-k},
\end{eqnarray}
where
\begin{eqnarray}
    \xi_k &=& \text{sgn}(g_{\alpha}(k)) \sqrt{\frac{-\mu - 2J f_{\alpha}(k) + E_k}{2E_k}}, \nonumber \\
    \eta_k &=&- \frac{|g_{\alpha}(k)| \sqrt{\Delta^2- \delta^2}}{\sqrt{2E_k(E_k - \mu - 2J f_{\alpha}(k))}}.
\end{eqnarray}
Thus, the model can be mapped to the BdG form through the complex Bogoliubov transformation and the ground state is
\begin{eqnarray}
|g\rangle\rangle=\Pi_k |g_k\rangle\rangle=\Pi_k (u_k + v_kc^{\dagger}_kc^{\dagger}_{-k})|0\rangle\rangle\label{sm1}
\end{eqnarray}
and
\begin{eqnarray}
|\bar{g}\rangle=\Pi_k |\bar{g}_k\rangle=\Pi_k (\bar{u}^*_k + v^*_kc^{\dagger}_kc^{\dagger}_{-k})|0\rangle\label{sm2}
\end{eqnarray}
with $\langle \bar{g}|g\rangle\rangle=1.$ Here $v_k=\eta_k, $ ${u}_k=i\sqrt{\frac{\Delta-\delta}{\Delta+\delta}} \xi_k,$ and $\bar{u}_k=-i\sqrt{\frac{\Delta+\delta}{\Delta-\delta}} \xi_k.$

\section{Symmetry Behavior}\label{appa}

The non-Hermitian model adheres to time-reversal symmetry (TRS, \(\mathcal{T}\)), particle-hole symmetry (PHS, \(\mathcal{C}\)), and chiral symmetry (CS, \(\Gamma\)). Specifically in momentum space, 
these symmetries are defined by:
\begin{itemize}
    \item \(\mathcal{T} H^*(k) \mathcal{T}^{-1} = H(-k)\) with \(\mathcal{T}\mathcal{T}^* = \pm 1\) and \(\mathcal{T} = \mathcal{U}\mathcal{K}\), where \(\mathcal{U}\) is a unitary operator.
    \item \(\mathcal{C} H^T(k) \mathcal{C}^{-1} = -H(-k)\) with \(\mathcal{C}\mathcal{C}^* = \pm 1\) and \(\mathcal{C} = \mathcal{U}\mathcal{K}\).
    \item \(\Gamma H^{\dagger}(k) \Gamma^{-1} = -H(-k)\) with \(\Gamma = \mathcal{T}\mathcal{C}\).
\end{itemize}

With $H$ being in BdG structure. Here, \(\mathcal{T}\) is an anti-unitary operator represented by complex conjugation \(\mathcal{K}\). This framework ensures the existence of eigenvalues \(E\) and \(E^*\) corresponding to each eigenstate \(|\psi\rangle\) and \(|\psi^*\rangle\). The PHS is also anti-unitary, commuting with the Hamiltonian to guarantee that eigenvalues occur in pairs \(\pm E\). Furthermore, the model exhibits parity-time (PT) symmetry, which maintains both spatial and temporal symmetry. The introduction of long-range (LR) effects does not alter these symmetry properties.

\textbf{PT Symmetry:} PT symmetry represents a combination of parity and TRS, ensuring space-time inversion of the Hamiltonian, which significantly impacts the physical properties of the system. In this context, the model adheres to both TRS and PHS, resulting in the occurrence of eigenvalues \(E(E^*)\) and \(E(-E)\), respectively. Notably, the current non-Hermitian Hamiltonian yields real eigenvalues, classifying it as a pseudo-Hermitian Hamiltonian, which satisfies the condition:
\begin{equation}
    H_{NH}^{\dagger}(k) = \eta H_{NH} \eta^{-1},
\end{equation}
where \(\eta = U^{\dagger} \eta U\) and 
\[
    U = \frac{1}{\sqrt{2}} \begin{pmatrix} 
    1 & -1 \\ 
    1 & 1 
    \end{pmatrix}, \quad \sigma_z = \begin{pmatrix} 
    1 & 0 \\ 
    0 & -1 
    \end{pmatrix}.
\]
Thus, our Hamiltonian behaves analogously to a Hermitian Hamiltonian and does not exhibit skin effects, although this conclusion is system-dependent. This Hamiltonian falls into the pseudo-skew-symmetric category, with eigenvalues forming pairs \(\pm E\) and \(E, -E^*\), which leads to the absence of non-Hermitian skin effects.
\begin{figure}[H]
	\centering
\includegraphics[width=\columnwidth,height=8cm]{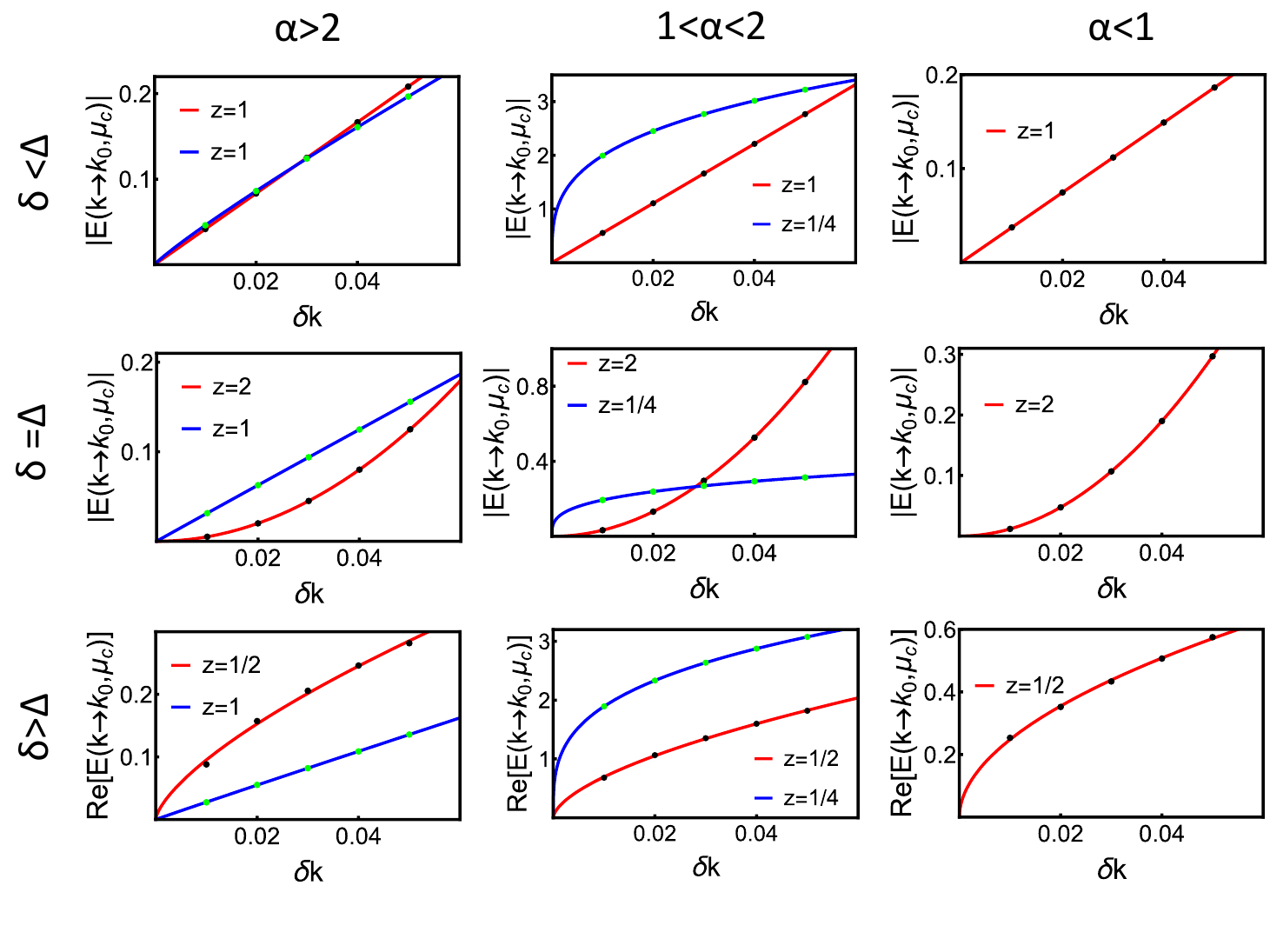}
	\caption{(Color online) Dynamical critical exponent for different ranges of the long-range decay parameter $\alpha$ and non-Hermitian parameter $\delta$. The blue and red colors correspond to criticalities $k_0=0$ and $\pi$ at $\mu_c=-2J\text{Li}_{\alpha}(\pm1)$ 
 with $J=1$ respectively.}
	\label{f2}
\end{figure} 
The Hamiltonian obeys pseudo-Hermiticity. For $(-\mu - 2J f_{\alpha}(k))^2 + 4(\Delta^2 - \delta^2) g^2_{\alpha}(k)\geq0$, and the eigenvalues $E_k$ are real-valued similar to a Hermitian Hamiltonian.This corresponds to a pseudo-Hermitian region. For $(-\mu - 2J f_{\alpha}(k))^2 + 4(\Delta^2 - \delta^2) g^2_{\alpha}(k)<0$, the time-reversal symmetry is broken and there occur many critical values $k_c$ satisfying
     $f_{\alpha}(k_c) =\frac{-4\mu J \pm \sqrt{(4\mu J)^2 - 4(4J^2 - 4(\Delta^2-\delta^2))(\mu^2 + 4(\Delta^2-\delta^2))}}{2(4J^2 - 4(\Delta^2-\delta^2))}.$
The Hamiltonian forms a Jordan block as
  $sh_{k_c}s^{-1}=\begin{pmatrix}
0 & 1 \\
0 & 0
\end{pmatrix}$  ,
with 
   $s= \begin{pmatrix}
1 & 0 \\
-\mu-2Jf_{\alpha}(k_c) & -i\Delta+\delta g_{\alpha}(k_c)
\end{pmatrix}$. This leads to the emergence of coalescing ground states by coalescence of two eigenvectors of the Jordan block. Thus, it represents the coalescing region. 

The Hamiltonian gives energy dispersion (as in Eq. 4 of the main text)
with criticalities at \(k_0 =(0, \pi) \) corresponding to \(\mu_c=-2J \text{Li}_{\alpha}(\pm 1)\). Here the criticality at $k_0=0$ is sensitive towards long-range effects and at $k=\pi$ is sensitive towards non-Hermitian effects (Fig. \ref{f2}). 

\section{Scaling of the Quantum Geometric Tensor and Fidelity Susceptibility}
For a Hermitian Hamiltonian $H(X)$ with $X = \{X^{n}\}_{n=1}^N$, the quantum distance on the ground state manifold is defined as:
\begin{eqnarray}
    F(X + dX) &=& |\langle \psi(X) | \psi(X + dX) \rangle| \nonumber \\
    &=& 1 - \frac{1}{2} (dX)^2 \chi_a + \ldots
\end{eqnarray}
with $0 < F(X + dX) < 1$. This is called fidelity, and under the $dX \rightarrow 0$ condition, an expansion yields:
\begin{eqnarray}
    g_{aa^{\prime}} = \frac{\partial X_{a}}{\partial X} \frac{\partial X_{a^{\prime}}}{\partial X} \chi_{aa^{\prime}},
\end{eqnarray}
where $g_{aa^{\prime}}$ is the fidelity susceptibility.

Here, $dX$ is the short displacement, $dX_{aa^{\prime}}$ is the direction of displacement, and $\chi_{aa^{\prime}}$ is the quantum geometric tensor in $aa^{\prime}$ parameter space, which is defined as:
\begin{equation}
    \chi_{aa^{\prime}} = \frac{1}{2} \left( \langle \partial_{a} \psi | \partial_{a^{\prime}} \psi \rangle - \langle \partial_{a} \psi | \psi \rangle \langle \psi | \partial_{a^{\prime}} \psi \rangle \right),
\end{equation}
with $\psi$ as the wavefunction associated with the BdG structure.
The quantum metric tensor and Berry curvature, which are the symmetric and antisymmetric parts of the quantum geometric tensor, frame a complete picture as:
\begin{eqnarray}
    g_{aa^{\prime}} &=& \text{Re}[\chi_{aa^{\prime}}] = \frac{1}{2}(\chi_{aa^{\prime}} + \chi_{a^{\prime}a}), \nonumber \\
    F_{aa^{\prime}} &=& \frac{1}{2}i(\chi_{aa^{\prime}} - \chi_{a^{\prime}a}) = \partial_{a} A_{a^{\prime}} - \partial_{a^{\prime}} A_{a},
\end{eqnarray}
where $A_{a} = \langle \psi | \partial_{a} \psi \rangle\rangle$.

The non-Hermitian adaptation of the above differs significantly. As non-Hermitian systems are spanned in a bi-orthonormal basis, we need to reformulate the quantum geometric tensor similarly:
\begin{eqnarray}\label{qgt}
    g_{aa^{\prime}} &=& \left(\langle \partial_{a} \psi | \partial_{a^{\prime}} \psi\rangle \rangle - \langle \partial_{a} \psi | \psi\rangle \rangle \langle \psi | \partial_{a^{\prime}} \psi\rangle \rangle \right)\nonumber\\
    &+& a \Leftrightarrow a^{\prime},
\end{eqnarray}
with $\langle \psi | \psi\rangle \rangle = 1$.

This raises a question about symmetry. In Hermitian systems, the real and symmetric aspects coincide, and the quantum geometric tensor is defined as the symmetric part. However, in the non-Hermitian context, ambiguity arises: should we consider the real part, the symmetric part, or both? The available literature on this topic is limited, and a convenient approach is required. Refs.~\cite{jiang2018topological,liao2021experimental} consider the real part of the quantum geometric tensor, while Refs.~\cite{zhu2021band,zhang2019quantum} prefer the symmetric part. We agree with the arguments of Ref.~\cite{chen2024quantum}, which incorporate the necessary conditions. Since the non-Hermitian quantum geometric tensor is spanned using a bi-orthonormal framework, complex eigenvalues naturally lead to complex geometry. The primary condition for the quantum metric tensor is symmetry around the dimensional index, while the real part may become an additional constraint. Thus, the approach depends on the geometry and becomes difficult to generalize across systems. In our case, the matrix is pseudo-skew-Hermitian, where the eigenvalues are either real with positive-negative pairs ($\pm E$) or complex conjugates with opposite signs ($E, -E^*$). Hence, the model behaves similarly to Hermitian Hamiltonians. Therefore, we do not pursue a foundational study of the quantum geometric tensor but consider the real and symmetric part of the quantum geometric tensor, which is convenient for our model and produces the entire phase diagram.
We write the above Hamiltonian in the Pauli spin basis ($\sigma_{x,y,z}$) with a 3D complex vector field, i.e.,
\begin{eqnarray}
    H(k) = \chi_{x,y,z}(k) \cdot \sigma_{x,y,z}\label{bdg1}
\end{eqnarray}
where
\begin{eqnarray}
    \chi_x &=& \frac{-i}{2} (2\delta) \sum_{l=1}^{L-1} \frac{\sin(kl)}{l^{\alpha}}, \nonumber \\
    \chi_y &=& \frac{1}{2} (2\Delta) \sum_{l=1}^{L-1} \frac{\sin(kl)}{l^{\alpha}}, \nonumber \\
    \chi_z &=& -\mu - 2J \sum_{l=1}^{L-1} \frac{\cos(kl)}{l^{\alpha}},\label{bdg2}
\end{eqnarray}
where each component of the vector $\chi$ is a function of $k$. In the symmetry-unbroken regime, we can construct the eigenstates of the matrix through the biorthonormal vectors as
\begin{eqnarray}\label{bdg3}
    |\psi_{+}\rangle &=& \left( \begin{matrix}
        \cos\frac{\theta}{2} e^{-i\phi} \\
        \sin\frac{\theta}{2} \\
    \end{matrix} \right),
    \hspace{0.2cm}
    |\psi_{-}\rangle = \left( \begin{matrix}
        \sin\frac{\theta}{2} e^{-i\phi} \\
        -\cos\frac{\theta}{2} \\
    \end{matrix} \right)
    \nonumber \\
    |\psi_{+}\rangle\rangle &=& \left( \begin{matrix}
        \cos\frac{\theta}{2} e^{i\phi} \\
        \sin\frac{\theta}{2} \\
    \end{matrix} \right)^*,
    \hspace{0.2cm}
    |\psi_{-}\rangle\rangle = \left( \begin{matrix}
        \sin\frac{\theta}{2} e^{i\phi} \\
        -\cos\frac{\theta}{2} \\
    \end{matrix} \right)^*
\end{eqnarray}
which is in the polar coordinate system with $\chi = r (\cos(\theta), \sin(\theta)\cos(\phi), \sin(\theta)\sin(\phi))$, where
\begin{eqnarray}
    r &=& \sqrt{\left(-\mu - 2J f_{\alpha}\right)^2 + (\Delta^2 - \gamma^2) \left( g_{\alpha}\right)^2} \nonumber \\
    \cos(\theta) &=& \frac{-i}{2r} (2 \delta) \sum_{l=1}^{L-1} \frac{\sin(kl)}{l^{\alpha}} \nonumber \\
    \tan(\phi) &=& \frac{(2\Delta) \sum_{l=1}^{L-1} \frac{\sin(kl)}{l^{\alpha}}}{2 \left(\mu - 2J \sum_{l=1}^{L-1} \frac{\cos(kl)}{l^{\alpha}}\right)}.\label{bdg4}
\end{eqnarray}
The biorthonormal vectors satisfy $\langle \psi_{b^{\prime}} | \psi_{b}\rangle \rangle = \delta_{bb^{\prime}}$ with $b = \pm$ and $\sum |\psi_{b^{\prime}}\rangle\rangle \langle\psi_{b}| = 1$. Generally, these relations hold except at exceptional points.
One can show that both $|\psi_+\rangle(|\psi_+\rangle\rangle)$ and $|\psi_-\rangle(|\psi_-\rangle\rangle)$ lead to the same Zak phase in the symmetry unbroken region.



 According to the definition, the quantum geometric tensor is expressed by Eq.~\ref{qgt}, and the real/symmetric part gives the quantum metric tensor, which also follows the structure of fidelity susceptibility.

\begin{eqnarray}
    g_{kk} &=& \frac{1}{2} \partial_k \hat{\chi} \cdot \frac{1}{2} \partial_k \hat{\chi} \nonumber \\
    &=& \frac{1}{4} (\hat{\chi} \times \partial_k \hat{\chi})^2 \nonumber \\
    &=& \left(\langle\psi_-| i \partial_{\phi} |\psi_-\rangle\rangle \partial_k \phi + \langle\psi_-| i \partial_{\theta} |\psi_-\rangle\rangle \partial_k \theta\right)^2 \nonumber \\
    &=& \left(\langle\psi_-| i \partial_k |\psi_-\rangle\rangle\right)^2
\end{eqnarray}
In this case, it is important to remember that the fidelity susceptibility is a function of $k$, and $\frac{\partial_k\hat{\chi}}{2}$ plays the role of a vielbein \cite{panahiyan2020fidelity}. 

Here the vectors $|\psi_-\rangle\rangle$ and $|\psi_-\rangle$ represent the biorthonormal eigenstates corresponding to the $-E_k$ eigenenergy of a BdG Hamiltonian. This state can be directly connected with the BCS ground states through Bogoliubov quasiparticles. The BdG Hamiltonian and its eigenstates can be written as in Eq.~\ref{bdg1}-\ref{bdg4}. Then we define the Bogoliubov quasiparticle operator through Eq.~\ref{qp} and the many-body ground state $\ket{g}$ is defined as the vacuum of these operators:
\begin{equation}
A_k |g\rangle\rangle
= 0.
\end{equation}

From the eigenvector $|\psi_-\rangle\rangle$, we identify:
\begin{equation}
u_k = -\cos\frac{\theta}{2}, \qquad v_k = \sin\frac{\theta}{2} e^{i\phi}.
\end{equation}

Then the BCS ground state is given by:
\begin{align}
|g\rangle\rangle &= \prod_k \left( u_k + v_k c_k^\dagger c_{-k}^\dagger \right) |0\rangle\rangle \\
&= \prod_k \left( -\cos\frac{\theta}{2} + \sin\frac{\theta}{2} e^{i\phi} c_k^\dagger c_{-k}^\dagger \right) |0\rangle\rangle,
\end{align}
and $\partial_{a} |g_k\rangle\rangle=\frac{\partial_{a}\theta}{2}(\sin\frac{\theta}{2} +\cos\frac{\theta}{2}e^{i\phi} c^{\dagger}_kc^{\dagger}_{-k})|0\rangle\rangle+( i\partial_{a}\phi \sin\frac{\theta}{2}e^{i\phi} c^{\dagger}_kc^{\dagger}_{-k})|0\rangle\rangle.$  Thus defining fidelity susceptibility using biorthonormal states $|g\rangle$ and $|g\rangle\rangle$ leads to the relation
\begin{eqnarray}
   g_{kk}  &=& \left(\langle\bar{g}| i \partial_k |g\rangle\rangle\right)^2.
\end{eqnarray}
Note that the fidelity susceptibility scales as square of extended
Berry connection in the ground states. Also the fidelity
susceptibility is a gauge dependent quantity which
extended Berry connection still remain gauge dependent.

\end{document}